# Gender Balance and Inclusion Within The Australian Space Community: An Overview Of Delegates At The 2018 Australian Space Research Conference

Eriita G. Jones [1], Jonathan Horner [2], Ann Cairns [3] and Wayne Short [4]

[1] School of IT and Mathematical Sciences, University of South Australia, Adelaide, SA, Australia.
[2] University of Southern Queensland, Centre for Astrophysics, Toowoomba, Queensland 4350, Australia.
[3] Sydney Observatory, Museum of Applied Arts and Sciences, New South Wales, Australia.
[4] National Space Society of Australia Ltd, GPO Box 7048, Sydney, New South Wales, 2001, Australia.

**Summary:** A significant gender disparity is widely known to occur within the fields of science, technology, engineering and mathematics (STEM). Women are both under-represented in their participation in Australian STEM education (compared to their fraction within the population at large), and face a much higher attrition rate from STEM subjects throughout secondary and tertiary education. Even within STEM careers the proportion of women frequently generally decreases with increasing position of seniority. In 2015 organisers and members of the Australian Space Research Community began to analyse the data from participants of the Australian Space Research Conference, in order to derive statistics relating to the gender balance of the conference. In this paper, the data from the most recent conference (held on the Gold Coast, in September 2018) is analysed, considering the gender demographics of delegates, presenters (within numerous categories), and awards. This year, we also present the dimensions of career type and academic level of conference attendees. The resulting trends are compared to those of another national space research conferences – the Scientific Meeting of the Astronomical Society of Australia. The results show that the representation of females dropped to the lowest levels since formal data collection began in 2015, with females comprising 23% of conference attendees, and 21% of all session presentations (posters and talks combined). Gender parity was achieved however amongst the invited Plenary presentations. Recommendations for removing some of the barriers to female participation in the conference and encouraging greater gender balance and diversity in future are discussed.

## Introduction

The decreasing proportion of women with increasing qualification and career level – or so called 'leaky pipeline' effect – is well evidenced within STEM careers [1]. Despite representing ~50.0% of the Australian population (as of 2017), and more than 55% of the commencing university students ([2]; as of 2017), a report into Australia's STEM workforce released in 2016 (using 2011 data) found that women represented only 29% of STEM-qualified university graduates. The proportion of women within STEM careers lay even lower at 16.5% [3].

Although arguments are frequently made to the contrary, statistical analyses show that the leaky pipeline effect does not in itself justify nor explain the under-representation of women at senior levels in STEM fields. For example, [4] conducted a broad examination of the participation of women in NASA's planetary science program spanning 1975-2009, demonstrating that the low numbers of women entering and remaining in the field of planetary sciences did not explain the numbers of women in science leadership. Even when accounting for their representation within the field, women in the planetary sciences were found to be significantly underutilised and lacking from leadership roles.

The causes behind this underrepresentation of women, and the many biases that women with STEM careers face, are largely beyond the scope of this paper. The barriers to female participation in STEM are known to be present very early, despite no innate cognitive gender differences. For example, the gender stereotyping of scientists (perceived to be male) is observed in the majority of children aged 9-11, and by this age girls are already statistically less confident in their ability to be successful at mathematics (compared to males) [5]. This broad gender socialization involving a perceived and influenced relationship between gender and STEM, and resulting observed gender disparities, is likely driven by a number of complex factors, including [1]:
- Gendered roles and expectations (e.g. primary child caregiver roles).
- Gendered associations with STEM fields.
- Skewed gender ratios (females outnumbered) leading to impaired sense of belonging ('imposter syndrome').
- Lack of female role models and mentors (e.g. female STEM teachers [6])
- Bias in the assessment of qualifications and provision of opportunities.

However, it is important to briefly mention why we should address the under-representation of women and minority groups in space research. Demonstrating that there are both (i) successful scientists from the full spectrum of gender, ethnic and cultural diversity, and that (ii) the achievements of these scientists are promoted and celebrated in equal measure to those from the dominantly represented groups (primarily Caucasian males), is essential for inspiring future space researchers to enter and persist within their field, and to maintain a diversity of thought. Consistent with the 'intersection feminist approach' – which aims to examine and understand how inequalities are often simultaneously experienced by minorities [7] – promoting a climate of inclusion for women in STEM can also promote the inclusion and removal of bias towards other minority groups. For example, women from the non-dominant cultural or ethnic groups [8]; and those who identify as LGBTQIA: lesbian, gay, bisexual, trans*[1], queer, intersex or asexual [9]. Confronting and removing the barriers to equity and inclusion is an issue not restricted to those of the female gender – it is a problem deserving attention from all members of society [10].

In response to these issues, there is now a strong push in many countries, and within many organisations, to raise awareness and address the marginalisation of the potential female workforce from STEM careers. Examples include:
- The Science in Gender Equity (SAGE) initiative of the Australian Academy of Science, which includes 45 Australian research organisations committed to adopting the Athena SWAN model for improving gender equity in STEM[2].

---

[1] Defined in the Oxford English Dictionary as: "originally used to include explicitly both transsexual and transgender, or (now usually) to indicate the inclusion of gender identities such as gender-fluid, agender, etc., alongside transsexual and transgender."
[2] https://www.sciencegenderequity.org.au/about-sage/

- The Astronomical Society of Australia's Pleiades Awards, which acknowledge astronomy groups that are taking appropriate steps to address equity and diversity, and are demonstrating a strong commitment to improving the work environment for women and the representation of women in a range of roles[3].
- There is a recognised lack of diversity in participation in NASA missions, particularly at senior levels. To address this, in late November 2018, NASA's Science Mission Directorate held a Diversity Workshop to discuss strategies to broaden and remove the barriers to participation in NASA missions. One key recommendation was to engage a more diverse team of mission principle investigators.

Conferences such as the Australian Space Research Conference (ASRC) cannot single-handedly fix the problems of gender equity and under-representation of females within STEM careers – nor the many flow-on effects of this issue. They cannot directly address the significant pay-disparity that those women with STEM careers are more than likely to encounter. The 2018 report into Australian graduate careers identified that female graduates earn considerably less than males in nearly every industry, with statistically significant graduate wage gaps occurring within STEM fields (Engineering being an exception), and that this pay disparity is also observed three years after graduation [11]. Nor can they directly address the negative impacts of workplace cultures dominated by males that females, and minority groups, are likely to experience in a STEM career. Such impacts were documented by a submission from industry group Professional Australia to the 2016 Senate inquiry into gender segregation in the workplace [12] (see also the Australian Human Rights Commission submission [13]; and numerous academic publications, e.g. [9], [14], [15]), and included a range of experiences such as 'a culture which rewarded long working hours, women professionals not being part of the "boy's club", women being subject to sexist remarks and the technical expertise of women being regarded less seriously than that of their male colleagues' [12].

Nonetheless, all scientific and technical meetings have the power to both inspire and support change. Through: showcasing the merits and achievements of women in STEM; highlighting to younger students the presence of female and non-binary (identifying as anything other than exclusively female or male) role models within the field with whom they can identify; holding events to support and nurture minority groups within STEM and to recognise the unique challenges and career hurdles they face; supplying a clear code of conduct to ensure that delegates are held accountable to behaving in an appropriate manner and providing a positive and supportive meeting environment (the ASRC's Code of Conduct can be found here[4]); and by providing a united voice that can speak to government and the media on these issues (e.g. by conference organisers and participants demonstrating through public media their commitment to advancing gender equity [16], [17]),– conferences such as the ASRC have an important role to play in striving for equity and inclusion. For this reason, members of the ASRC recognised in 2015 the need to collect statistics of the gender demographics of conference organisers, participants, attendees, and merit recipients. This data set, which is accumulated annually, allows researchers to examine the changing state of the national space research community, and to provide a metric against which the success of future initiatives can be measured. Two prior publications presented an analysis of the gender balance of the 2015 and 2016 Australian Space Research Conferences [18], [19]. In this paper, the results from the 18th Australian Space Research Conference (2018) are presented and compared to the years prior, recommendations from the lunchtime 'Women In Space' meeting are discussed, and potential future directions for the conference are suggested.

---

[3] https://asa-idea.org/the-pleiades-awards/
[4] http://www.nssa.com.au/18asrc/CONDUCT/

# Gender Demographics of Delegates

The 18th Australian Space Research Conference, held on the 24th – 26th of September 2018 in the Gold Coast, Queensland, is the only national conference that brings together researchers, technicians, policy makers, educators and students from across all disciplines of space. The 2018 meeting combined 187 delegates and featured three parallel presentation streams spanning Earth Observation, Space Physics, Space Engineering, Aerospace Medicine, Space Business and Industry, Space Entrepreneurs, Space Law, Space Missions, Astrobiology, Mars, Education, Culture and History. The morning sessions were focussed on plenary presentations, and combined all conference participants in one room. Additional sessions throughout the conference included poster presentations, a town hall discussion, a public talk (Mars Society Australia's David Cooper Memorial Lecture), the annual Women In Space lunch, and the cocktail function and dinner.

In contrast to the gender demographics presented in the 2015 and 2016 proceedings publications [18], [19], the results obtained this year were derived from the participant's self-identification of gender as part of the online registration process[5]. During that process, delegates were presented with the gender options of 'Female', 'Male', 'Non-binary', or 'Prefer not to say'. All registrants selected one of these options. The sole exception to the clear-cut data provided by this process impacts our results for the gender demographics of the conference committee [20]. Members of the program or organising committee who did not attend the conference (and therefore did not use the online registration system) were assigned a gender for the purposes of this study using either the previous meeting data, or from an online search of their professional webpages (such as their institution profile, or LinkedIn profile). This affected 5 of the 27 committee members. We acknowledge that there is potential that we have mis-gendered these individuals in our dataset[6].

Table 1 presents the results obtained for the 18th Australian Space Research Conference (ASRC), with the data from the 16th conference for comparison (data from the 17th ASRC was not available). Although the number of delegates has increased from 2016 to 2018, the number of female delegates and hence the overall proportion decreased (by 3.4%). Females represented less than one quarter of the 2018 delegates. The plenary presentations were equally split between female and male presenters, by decision of the program committee. The representation of females on both the program and organising committee increased slightly, to ~19% and ~29% respectively.

*Table 1: The gender distribution across the 18th and 16th Australian Space Research Conferences.*

|  | 18th ASRC | | | | | 16th ASRC | | | |
|---|---|---|---|---|---|---|---|---|---|
|  | **Female** | **Male** | **Non-Binary** | **Non-disclosed** | **Total** | **Female** | **Male** | **Non-Binary** | **Total** |
| **Delegates** | 23.0% | 76.5% | 0% | 0.5% | 187 | 26.4% | 73.6% | 0% | 174 |
| **Talks** | 24.1% | 74.2% | 0% | 1.7% | 120[*] | 28.9% | 71.1% | 0% | 121 |

---

[5] The acquisition of our data in this manner was one of the key recommendations of the 2017 Women in Space lunchtime meeting, and was implemented to ensure that the risks of accidentally mis-gendering participants could be minimised.
[6] If you feel this may be the case, please contact the corresponding author.

| | | | | | | | | |
|---|---|---|---|---|---|---|---|---|
| **Posters** | 15.0% | 80.0% | 0% | 5.0% | 20[*] | 31.8% | 68.2% | 0% | 22 |
| **Plenaries** | 50.0% | 50.0% | 0% | 0% | 8 | 44.4% | 55.6% | 0% | 9 |
| **Student awards** | 37.5% | 62.5% | 0% | 0% | 8 | 33.3% | 66.7% | 0% | 6 |
| **Program committee** | 19.2% | 80.8% | 0% | 0% | 26 | 17.6% | 82.4% | 0% | 17 |
| **Organising committee** | 28.6% | 71.4% | 0% | 0% | 7 | 27.3% | 72.7% | 0% | 11 |

[*]These figures refer to the number of talks and posters presented. They do not correspond to the number of delegates presenting a talk or poster, as some delegates gave multiple presentations.

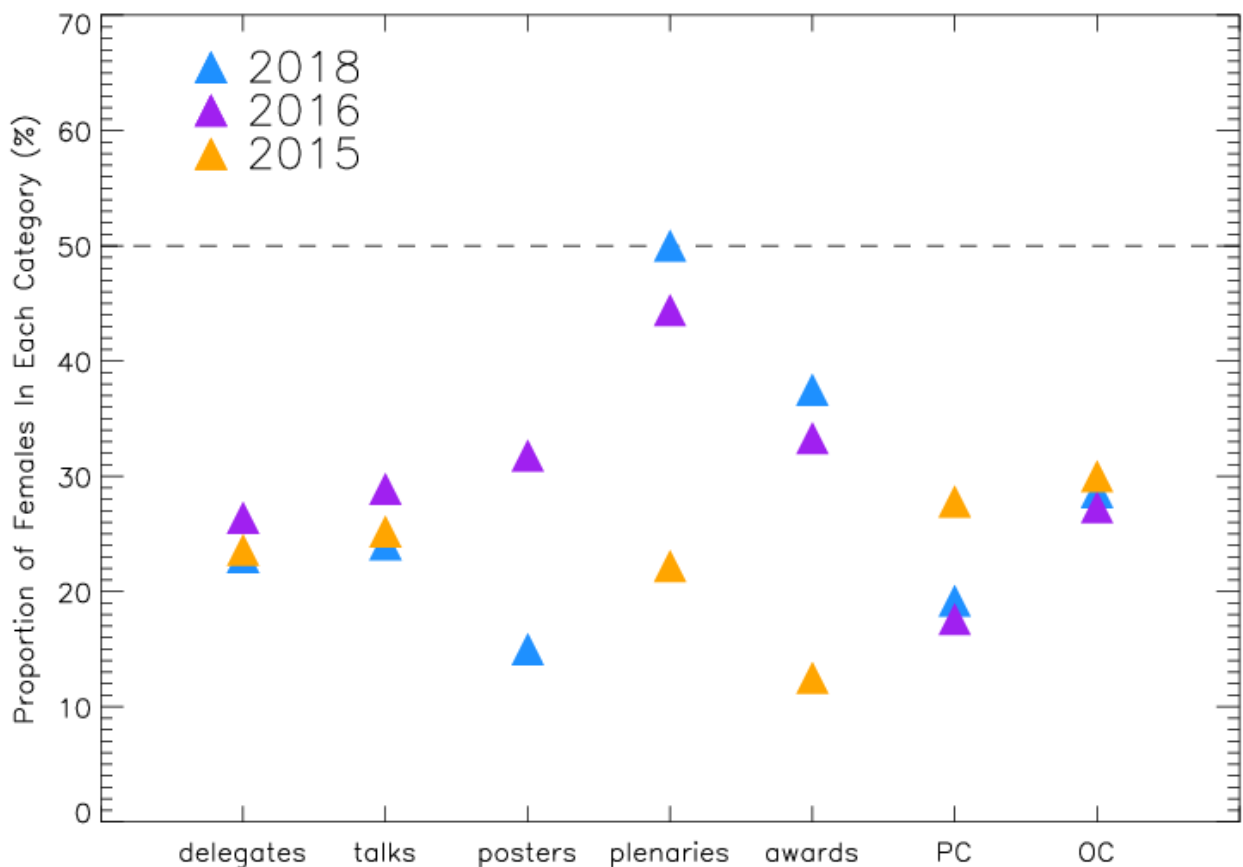

*Figure 1: The fraction of women within various categories at the Australian Space Research Conference (ASRC) in 2018, 2016 and 2015 (data for the ASRC in 2017 was not available). PC = Program Committee; OC = Organising Committee. The dashed line highlights the 50% fraction (gender parity) as a guide.*

The relationship between gender representation and career type and academic career stage was examined through utilisation of the registration type and position description provided during the registration process. Not all field entries could be used to identify career type, for example, '1 day registration - Early bird' and position 'Scientist' would be insufficient. Broad but disjoint categories of 'Student', 'Academic' and 'Industry/Government' could be identified however. For example, the category 'Academic' included registration types 'Academic registration', 'Academic registration - Normal', and 'Academic registration - Early bird', as well as a range of position descriptors, including 'Professor', 'Lecturer', etc. Two sub-categories of 'Academic' could also be identified, although few registrants provided sufficient information to assign them to these categories so the counts are low: Academic Tier 1 (including 'Lecturer', 'Senior Lecturer', 'Research Associate', 'Research Fellow', 'Postdoctoral Fellow'), and Academic Tier 2 (including 'Head of School', 'Professor', 'Associate Professor', 'Adjunct Professor'). The category 'Student' included student registration types as well as position descriptors such as 'PhD student' and 'PhD candidate'. The results are presented in Table 2. 137 of the 187 delegates (73%) could be assigned to a career type. The zero value non-binary gender column is removed for brevity. From Table 2, the available data indicate that the representation of females was highest within the academic field (at 25% female), although the proportion of females within the student and industry or government career categories was similar (23-24%). The representation of females within the earlier academic career stage or more junior career progression (Tier 1) was significantly higher than within the later career/more senior Tier 2 academic level (25.0% compared to 16.7%). It is however important to note that only 50% of delegates identified as having an academic career could be subdivided into Tier 1 and Tier 2 categories, so these statistics are indicative only and should not be over-interpreted.

Table 2: The gender distribution of delegates at the 18$^{th}$ ASRC by career type and academic career stage.

|  | Female | Male | Non-disclosed | Total (% delegates) |
|---|---|---|---|---|
| **Student** | 23.4% | 76.6% | 0% | 47 (25.1) |
| **Academic (all)** | 25.0% | 75.0% | 0% | 40 (21.4) |
| (Tier 1) | (25.0%) | (75.0) | 0% | (12) |
| (Tier 2) | (16.7%) | (83.3) | 0% | (6) |
| **Industry/Business/Government** | 24.0% | 76.0% | 0% | 50 (26.7) |
| **Other/Unknown** | 20.0% | 78.0% | 2.0% | 50 (26.7) |

## Community Feedback: Women In Space Meeting

In keeping with the previous four Australian Space Research Conferences, the 2018 conference held the fourth 'Women In Space' lunch meeting to examine and discuss questions of equity within the space community. All delegates were welcomed to the meeting (regardless of gender), and approximately thirty delegates attended. The meeting endeavoured to be a safe environment where all respectful inputs were welcomed, and in keeping with this no participants were identified in the meeting minutes, and no specific contributions to the meeting will be shared. The following general feedback from the meeting participants can be described:

- The establishment of a closed space community group for communication and discussion of issues relating to gender balance and diversity was suggested. The group would ideally have an online presence and visibility, and would represent diversity in career types, career level and age. The group could provide a voice to help shape the future space meetings and encourage fair representation.
- The continuing issues of gender bias were acknowledged (e.g. women forming a decreasing fraction of the workforce for roles of increasing seniority), and a desire to address gender bias through emphasis on merit and raising awareness was expressed. The need for the Australian Space Agency to also reflect the gender, ethnic and cultural diversity within space researchers was recognised.

- The importance of mentorship and female researchers having access to female mentors was discussed.
- A representative from the Australian Space Agency, Executive Director, Anntonette Dailey, spoke about the agency's emphasis on Women in STEM and focus on championing funding for this cause, and it was recognised that the space community could assist through the provision of data to the agency to support the agenda of women in space.

## Discussion: Space Industry Context

The 18th Australian Space Research Conference in 2018 was found to be largely male-dominated, with the male:female ratio amongst attendees being 3.3:1. The gender ratio amongst presenters (combining both talk and poster presentations) was skewed towards over-representation of male delegates, at 3.5:1. When considered separately, the ratios were 3.2:1 for speakers and 5:1 for poster presenters, showing considerable over-representation of male-led research in posters presenters(compared to the conference attendance ratio). The ratio amongst plenaries was 1:1. Female delegates were found more likely to be talk recipients. Out of the male delegates, 55% presented one or more talks, while 58% of female delegates presented one or more talks. Females were over-represented amongst student award recipients (compared to their representation as conference attendees), with the male:female award ratio being 1.7:1. An additional dimension to the analysis of the 2018 dataset was the examination of the gender balance with career type, which we hope to continue in our examination of future meetings. he male:female ratios amongst academics, students, and professionals within industry/government were found to be very similar to one another (averaging 3.1:1). When further sub-dividing the data into categories reflecting academic career stage, it was found that females formed a higher proportion of Tier 1 academic delegates than Tier 2 (male:female ratios of 3.0:1 and 5.0:1 respectively). This data suffers from small number statistics and may not be truly representative, however it does flag a potentially important means of monitoring the decreasing representation of women as one moves to the higher levels of the traditional STEM career progression in our future data collection efforts.

Compared to previous years, the gender balance amongst delegates was further from parity at this year's meeting. However, the balance amongst award recipients and plenaries has improved noticeably through the period of data collection. In contrast, the proportion of both oral and poster presentations contributed by female delegates was lower than in both 2016 and 2015. These statistics highlight the importance of implementing initiatives to try to remove, or at least mitigate, the barriers to female attendance and participation in the conference.

Another large scientific space-research gathering within Australia is the Annual Scientific Meeting (ASM) of the Astronomical Society of Australia (ASA). Although different from the ASRC in terms of the breadth of disciplines incorporated within the conference (members of the ASA predominately come from the astronomy, astrophysics, and astronomical instrumentation research fields) the ASA's ASM nevertheless provides a useful point of comparison. Figure 2 compares the gender statistics for the awarded talks at the ASA's ASM and ASRC over the past 3 years. The ASA's ASM has consistently shown a higher representation of women amongst the delegates, the scientific/program organising committee, and the awarded talks (see Table 3 and additional statistics in Appendix Table 2). Unlike for the ASRC, women formed a higher fraction of the ASA's ASM scientific/program organising committee than they did the logistics/organising committee. For both conferences, women typically had a similar or slightly higher representation amongst talk recipients than their proportion as conference attendees. Although there are many complex factors involved in the resulting gender distribution of these conferences, the Astronomical Society of Australia has, for a number of years, undertaken a number of significant initiatives to improve the diversity and equality within their scientific community (described below). It is likely that these initiatives are partially responsible for the higher-representation of women amongst the delegates and presenters at the ASA's ASM, compared to the ASRC.

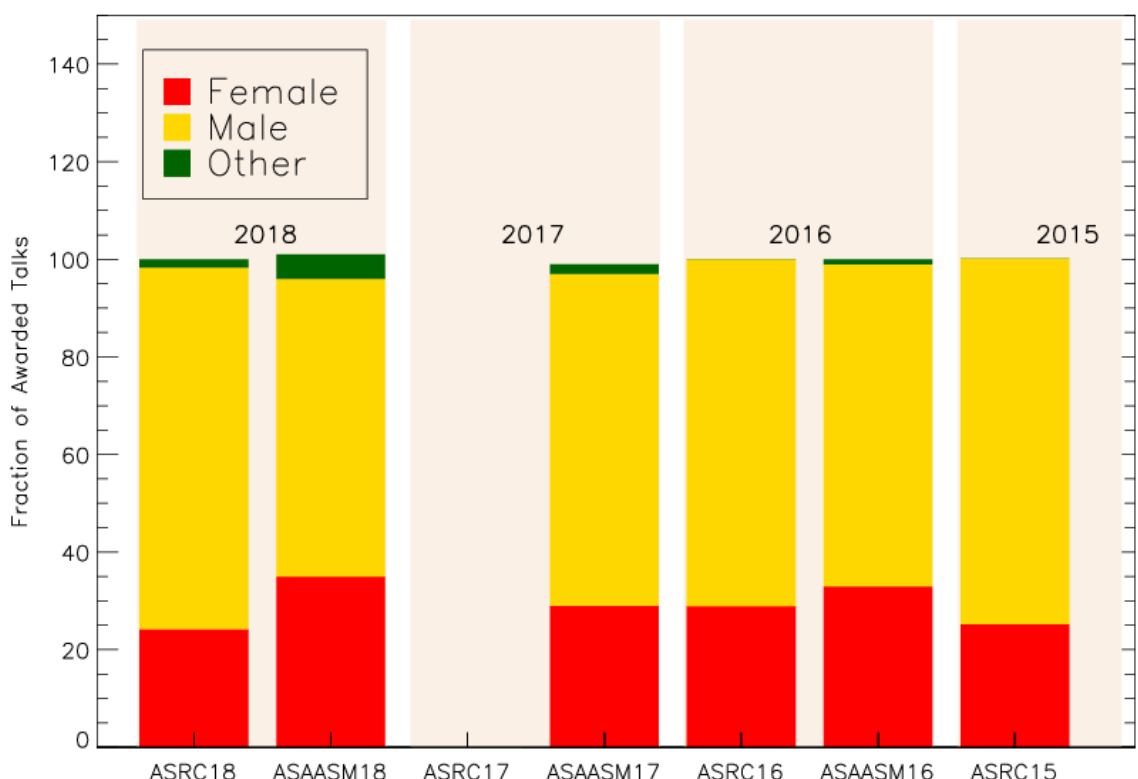

*Figure 2: Comparison between the gender demographics of awarded talks at the Australian Space Research Conference (ASRC) and the Annual Scientific Meeting of the Astronomical Society of Australia (ASA), between 2015-2018. Data for the ASRC in 2017 and the ASA's ASM in 2015 was not available. The 'Other' category here includes both participants who did not disclose their gender, and those who identified as being non-binary.*

*Table 3: The male:female ratio at the recent Australian Space Research Conferences and Annual Scientific Meetings of the Astronomical Society of Australia.*

|  | Australian Space Research Conference | | | Annual Scientific Meeting of the Astronomical Society of Australia | | |
| --- | --- | --- | --- | --- | --- | --- |
|  | 2018 | 2016 | 2015 | 2018 | 2017 | 2016 |
| **Delegates/Attendees** | 3.3:1 | 2.8:1 | 3.2:1 | 1.8:1 | 2.2:1 | 2.3:1 |
| **Awarded Talks*** | 3.1:1 | 2.6:1 | 3.0:1 | 1.7:1 | 2.3:1 | 2:1 |

*Calculated from the fraction of talks presented by males and females. This is not the same as the number of male and female delegates awarded talks as some delegates presented multiple talks.

# Progress And Future Directions

A number of scientific organisations and gatherings are taking a lead role in championing the cause for gender equity, and greater inclusion of gender, ethnic and cultural diversity within STEM fields. Examples include:
- The 'Women In Space' conference is organised by a number of scientists and engineers at US and Canadian institutions (and is an expansion of the "Women In Planetary Science" 2018 conference), aiming to promote the work and achievements of women and minorities in space and planetary science within academia and industry. The conference hosts panels focusing on the challenges faced by the LGBTQ+ community, those faced by women of colour, and harassment issues, within STEM. The inaugural conference held in 2018 had 100 attendees, out of which 85% identified as female, 2% as non-binary, and 13% as male (personal correspondence with conference organisers). The second conference was held in February 2019 at Arizona State University, whilst this paper was under review[7].
- The Astronomical Society of Australia established a 'Inclusion, Diversity, and Equity in Astronomy' (IDEA) Chapter in 2016 (an expansion of the original 'Women In Astronomy' Chapter, which was established in 2009). An associated inaugural 'Diversity in Astronomy' Workshop was held in 2017 to focus on issues of bullying, harassment, cultural sensitivity and diversity within Astronomy, and to highlight effective strategies for inspiring change. he IDEA Chapter offers a 'Conference Gender Balance' Endorsement for those gatherings that can demonstrate a focus on addressing issues of gender equity and diversity within STEM (for details see footnote[8]), and also organises the *Pleiades* awards, to recognise the work of Australia Astronomical research groups to promote and support equity and diversity.[9]

The Australian Space Research Conference organisers have recently taken a number of steps towards increasing diversity and gender equity. A new chapter to raise the profile of women in space sciences and industries is being established (as of October 2018), modelled on the successful ASA 'Women In Astronomy' Chapter mentioned above. The new ASRC Chapter is open to all genders, and all interested in joining and supporting the work of the chapter are encouraged to contact the conference organisers. In 2015, a policy was implemented to obtain a gender balance within the invited Plenary Speakers, with the same number of male and female speakers chosen, and for the first time in 2018 gender parity was achieved. Also in 2018, conference registrants were asked, if willing, to disclose their gender during the registration process, thereby enabling accurate gender statistics to be collected. A further step is the post-conference analysis of attendee demographics and the publication of those results (as was done in previous years – e.g. [18], [19]), providing a resource that can be accessed by all members of the space research community.

In addition to supporting the recommendations made in the 2016 proceedings publication [19], the authors would like to make the following recommendations to the ASRC organisers:
1. More detailed data collection in future, to more accurately determine the participation of under-represented groups, and the effectiveness of implemented changes aimed at improving conference diversity and gender equity. Future data collection would ideally include: information on delegate career stage (e.g. Early Career Researcher, Mid-Career Researcher); delegate ethnicity; the gender distribution of session chairs, statistics on the number of talk applicants versus number of talks awarded, registration types (e.g. fee paid; 1 day registrations versus full conference registrations), and the gender and career-stage distribution of audience questions during conference sessions. A necessary step prior to future data collection (and for management of already gathered data) must be the establishment of an ASRC data policy involving guidelines and procedures related to data curation, including: data privacy and de-identification, data storage and secure/restricted data access.
2. In addition to the submission of a proceedings publication on the diversity and gender equity of the conference, an online post-conference statistical summary report or series of infographics on participant statistics is encouraged. This would provide transparency and communicate to the space research community that the conference is committed to doing its part to lessen the barriers to participation and improve the gender equity and diversity at the gathering.
3. Gender parity be maintained in the selection of invited plenaries, and also established for the selection of invited keynote presenters and invited session chairs.

---
[7] http://www.womeninspacecon.com/
[8] https://asa-idea.org/resources-to-take-action/conference-gender-balance-endorsement-policy/
[9] https://asa-idea.org/the-pleiades-awards/

A number of further steps that the ASRC organisers are requested to consider, derived from [21], [22], are:
- To aim for greater gender balance and diversity within the Program Committee by inviting senior leading female researchers, or members of minority groups (including Indigenous researchers), to participate. In turn, this will help the committee to minimise subconscious bias[10] and help ensure balance within the selection of speakers.
- Establish a clear set of measurable goals for increasing diversity and equity within the ASRC. These may include: a quota of invited female speakers or talks awarded to female lead authors (e.g. striving for a gender balance of >40% female speakers); levels of financial or in-kind support for increasing female and minority group attendance (e.g. scholarships to fund a travel support person, conference child care arrangement, etc.); ensuring family rooms or non-gendered bathrooms are available at the conference venue; provision of a private, quiet room for prayer and reflection; providing information on which of the suggested conference hotel venues are family friendly, etc. These goals would ideally be publicly visible on the conference website.
- Solicit feedback from the attendees on how to increase and support diversity at the ASRC, and on how current progress is being perceived. This would be particularly valuable in identifying factors which lower the female attendance to the conference. As conference data collection relates only to those who registered and/or submitted an abstract, there is no current information on members of the space community who encountered barriers which prevented their attendance.
- Offer a mentorship program for first-time female and/or cultural and ethic minority group attendees, by pairing them with senior researchers. The mentor's role would be to provide support and encouragement in conference activities such as networking, and attending and contributing at talks. This would help encourage a more diverse early career conference delegation.

## Conclusions

The analysis of the gender balance within the 2018 Australian Space Research Conference carried out in this paper reveals that, at that conference, the distribution of both oral and poster presentations was more strongly skewed towards male presenters than was the case at the 2015 and 2016 meetings, with the representation of females dropping to the lowest levels since formal data collection began in 2015. Females comprised only 23% of conference attendees, and 23% of all session presentations (posters and talks combined). Conference organisers have demonstrated a focus on trying to increase gender equity and tackle the issue of under-representation of female researchers by ensuring gender parity in the selection of invited plenaries, expanding the collection of data for statistical analyses, and providing that data for analysis. At this year's conference a significant step was taken as a Chapter was established to raise the profile of female researchers within the space research community. Clearly the unsolicited scientific contributions at conferences should be awarded presentations through an unbiased merit based system regardless of gender. Conferences such as the ASRC cannot be expected to achieve gender parity amongst delegates until the many issues of female under-representation throughout STEM education and careers have been addressed. Nevertheless, all scientific gatherings have a large contribution to make in encouraging and supporting the contributions of female and minority groups, and endeavouring to minimise the significant barriers to entry that these groups uniquely experience. Furthermore, in the words of the "Astronomy In Colour" initiative, studies such as this demonstrate a commitment to "understand[ing] the past and present repercussions of systemic oppression of marginalized groups on our ability to study the Universe."[11]

---
[10] The identification and minimisation of subconscious bias could also be addressed by encouraging all members of the program committee to take the Harvard Implicit Association test prior to assessing abstracts.

## Acknowledgements

The authors wish to extend their gratitude to the organisers of the Australian Space research Conference for continuing to supply data and support this research. We wish to offer our heartfelt thanks to the Astronomical Society of Australia for supplying three years of statistical data and reports. Thank you also to the Women In Space conference organisers for data supplied.

---

[11] The "Astronomy In Color" initiative is run by members of the astronomy community committed to increasing diversity by recognizing, confronting and removing the barriers to racial equity and inclusion. The initiative involves regular blog posts which promote the community and their astronomical research and highlight social justice issues http://astronomyincolor.blogspot.com/.

# Appendix

*Table A1: The gender distribution across the 18th and 16th and 15th Australian Space Research Conferences.*

|  | 18th ASRC ||||| 16th ASRC |||| 15th ASRC ||||
|---|---|---|---|---|---|---|---|---|---|---|---|---|---|
|  | Female | Male | Non-Binary | Non-disclosed | Total | Female | Male | Non-Binary | Total | Female | Male | Non-Binary | Total |
| **Delegates** | 23.0% | 76.5% | 0% | 0.5% | 187 | 26.4% | 73.6% | 0% | 174 | 23.6% | 75.9% | 0.5% | 191 |
| **Talks** | 24.1% | 74.2% | 0% | 1.7% | 120 | 28.9% | 71.1% | 0% | 121 | 25.2% | 74.8% | 0% | 129 |
| **Posters** | 15.0% | 80.0% | 0% | 5.0% | 20 | 31.8% | 68.2% | 0% | 22 | 0% | 92.3% | 7.7% | 13 |
| **Plenaries** | 50.0% | 50.0% | 0% | 0% | 8 | 44.4% | 55.6% | 0% | 9 | 22.2% | 77.8% | 0% | 9 |
| **Student awards** | 37.5% | 62.5% | 0% | 0% | 8 | 33.3% | 66.7% | 0% | 6 | 12.5% | 87.5% | 0% | 8 |
| **Program committee** | 19.2% | 80.8% | 0% | 0% | 26 | 17.6% | 82.4% | 0% | 17 | 27.8% | 72.2% | 0% | 18 |
| **Organising committee** | 28.6% | 71.4% | 0% | 0% | 7 | 27.3% | 72.7% | 0% | 11 | 30% | 70% | 0% | 10 |

*Table A2: The gender distribution across the 2018, 2017, and 2016 Astronomical Society of Australia Annual Scientific Meetings.*

|  | 2018 ASAASM |||| 2017 ASAASM |||| 2016 ASAASM ||||
|---|---|---|---|---|---|---|---|---|---|---|---|---|
|  | Female | Male | Non-Binary / Non-disclosed | Total | Female | Male | Non-Binary / Non-disclosed | Total | Female | Male | Non-Binary / Non-disclosed | Total |
| **Delegates** | 35% | 62% | 3% | 304 | 31% | 67% | 2% | 220 | 30% | 70% | <1% | 265 |
| **Talks** | 35% | 61% | 4% | 152 | 29% | 68% | 2% | 143 | 33% | 66% | 1% | n/a |
| **Posters** | 29% | 68% | 3% | 80 | 34% | 66% | 0% | 39 | n/a | n/a | n/a | n/a |
| **Scientific organising committee** | 50% | 50% | 0% | 12 | 63% | 37% | 0% | 16 | 47% | 47% | 6% | 17 |
| **Logistics organising committee** | 33% | 67% | 0% | 15 | 29% | 71% | 0% | 14 | 42% | 58% | 0% | 24 |